\documentstyle[12pt]{article}
\begin{document}
\def\Ket#1{||#1 \rangle}
\def\Bra#1{\langle #1||}
\def\ua{u_{t_1}}
\def\ub{u_{t_2}}
\def\uc{u_{t_3}}
\def\ud{u_{t_4}}
\def\va{v_{t_1}}
\def\vb{v_{t_2}}
\def\vc{v_{t_3}}
\def\vd{v_{t_4}}
\def\ut{u_t}
\def\vt{v_t}
\def\utt{u_{t'}}
\def\vtt{v_{t'}}
\def\tt{t'}
\def\up{u_{p}}
\def\vp{v_{p}}
\def\un{u_{n}}
\def\vn{v_{n}}
\def\upp{u_{p'}}
\def\vpp{v_{p'}}
\def\unn{u_{n'}}
\def\vnn{v_{n'}}
\def\omt{\Omega_t}
\def\omtt{\Omega_{t'}}
\def\omp{\Omega_p}
\def\ompp{\Omega_{p'}}
\def\omn{\Omega_n}
\def\omnn{\Omega_{n'}}
\def\Nt{{\cal N}_t}
\def\Ntt{{\cal N}_{t'}}
\def\Np{{\cal N}_p}
\def\Npp{{\cal N}_{p'}}
\def\Nn{{\cal N}_n}
\def\Nnn{{\cal N}_{n'}}
\def\endauthors{}
\def\authors#1\endauthors{#1}
\def\be{\begin{equation}}
\def\ee{\end{equation}}
\def\br{\begin{eqnarray}}
\def\er{\end{eqnarray}}
\def\bc{\begin{center}}
\def\ec{\end{center}}
\def\N {{{\cal N}}}
\def\A {{{\cal A}}}
\def\F {{{\cal F}}}
\def\B {{{\cal B}}}
\def\X {{{\cal X}}}
\def\Y {{{\cal Y}}}
\def\Z {{{\cal Z}}}
\def\0{\tilde{0}}
\def\ket#1{|#1 \rangle}
\def\bra#1{\langle #1|}
\def\adt{\alpha^{\dagger}_{\bf t}} 
\def\at{\alpha_{\bar{\bf t}}} 
\def\aadt{\alpha^{\dagger}_t} 
\def\aadtt{\alpha^{\dagger}_{t'}} 
\def\aat{\alpha_{\bar t}} 
\def\aatt{\alpha_{\bar {t'}}} 
\def\tb{{\bf t}} 
\def\pb{{\bf p}} 
\def\nb{{\bf n}} 
\def\e{\epsilon}
\def\sss{\scriptscriptstyle}
%
\title{Limitations of the number selfconsistent Random Phase Approximation}
\author{Alejandro Mariano$^a$\thanks{Fellow of the CONICET, Argentina.
On leave of absence from Departamento de F\'\i sica, Facultad de Ciencias
Exactas,  Universidad Nacional de La Plata, C. C. 67, 1900 La Plata,
Argentina.} 
 and Jorge G. Hirsch$^b$\\
{\small\it $^a$Departamento  de F\'\i sica, Centro de Investigaci\'on
y  de Estudios Avanzados del IPN, }\\
{\small\it A. P. 14$-$740 M\'exico 07000 D.F.}\\
{\small\it $^b$Instituto de Ciencias Nucleares, Universidad Nacional
Aut\'onoma de M\'exico, }\\
{\small\it A.P. 70-543 M\'exico 04510 D.F.}
}
\maketitle

\begin{abstract}
The Quasiparticle Random Phase Approximation equations are solved
taking into account the Pauli Principle at the expectation value level,
and  allowing changes in the mean field occupation numbers to minimize 
the energy while having the correct number of particles in the
correlated vacuum. 
The study of Fermi pn excitations in $^{76}$Ge using a realistic Hilbert
space shows that the pairing energy gaps in the modified mean field are
diminished up to one half of the experimental value when strong
proton-neutron correlations are present. Additionally, the Ikeda sum rule
for  Fermi transitions is violated due to the lack of scattering terms 
in the phonon operators. 
These results call for a critical revision of the double beta decay
half-lives estimated using the QRPA extensions when standard QRPA
calculations collapse.

\end{abstract}

\section{Introduction}

The random phase approximation (RPA) and its quasiparticle
generalization (QRPA) have been widely used in the last decades to study
electromagnetic transitions and beta decays in medium and heavy nuclei
\cite{Row70,Rin80}.
The proton-neutron quasiparticle random phase approximation (pn-QRPA)
has been extensively employed in the description of single and double
beta decays in vibrational nuclei. However the RPA develops a
collapse, i.e. it presents imaginary  eigenvalues for strengths beyond a
critical value of the force \cite{Vog86,Civ87,Eng88,Hir90a}.
 
A whole family of extensions of the QRPA, called renormalized QRPA
(RQRPA) are known that do not develop any collapse by implementing the
Pauli principle in a consistent way, beyond the simplest
 quasiboson approximation  \cite{Row68,Cat94,Toi95,Sch96,Duk96,Mut97,Sam99}.
However,
in its simplest versions there is a violation of the non energy weighted
Ikeda sum rule \cite{Hir96}.
Calculations to determine the amount of the violation and some
improvements to the RQRPA, in order to restore the sum rule,
have been presented \cite{Krm96a}. It has been shown that
treating simultaneously BCS and QRPA equations one can fulfil the Ikeda
sum rule for the Fermi case when a schematic model is used \cite{Del97}.

In recent articles we studied the expectation
values of the particle and quasiparticle numbers, the particle
number fluctuations  and the number of particle pairs with J=0, T=1 and
J=1, T=0 in the ground state of $^{76}$Ge as a function of the residual
proton-neutron interaction, using realistic Hilbert spaces
\cite{Mar98a,Mar98b}.
We found an important amount of particle number fluctuations in the RQRPA
ground state beyond the QRPA collapse, pointing out a source of
uncertainty in the RQRPA results. The analysis of the number of pairs
showed that the isoscalar-isovector phase transition found in exact
calculations is causing the QRPA collapse and is missed in the RQRPA
formalism.

In the present work we go a step further, studying $^{76}$Ge with a
renormalized RPA where at the same time the mean field is changed by
minimizing the energy and fixing the number of particles in the
correlated ground state.
While particle number fluctuations are smaller than in the previous
cases, they still exhibit a clear increase after the point of collapse.
More remarkably, the pairing gap is strongly reduced in comparison with
its experimental value, and the Ikeda sum rule is violated.
Both results cast serious doubts about the double beta decay half-lives
estimated using the QRPA extensions, in particular for those
nuclei where standard QRPA calculations collapse \cite{Toi95,Sch96,Mut97}.
Together with the conclusions obtained in \cite{Mar98a,Mar98b}, we get
a clear picture of the limitations associated with the QRPA extensions,
which at the end are the same common sense ask for: you cannot allow the
residual proton-neutron interaction to dictate the composition of the
ground state wave function without missing contact with actual nuclei,
even if the formalism allows you to overcome the collapse.

In Section 2 the renormalized gap and number equations are introduced,
whose relationship with the RQRPA equations is shown in Section 3. Some
relevant expectation values and the Ikeda sum rule are
discussed in Section 4, results for $^{76}$Ge are presented in Section 5
and the conclusions in Section 6.

\section {Renormalized Gap and Number Equations}
\label{sec1}

In this section the gap and number equations are obtained minimizing
the Hamiltonian expectation value $\bra{0} H \ket{0}$ in the correlated
vacuum $ \ket{0}$. The Hamiltonian is

\begin{equation}
H = H_p + H_n + H_{p, n}.\label{ham}
\end{equation}
The first two terms refer to the proton and neutron hamiltonians
\begin{equation}
H_t = \sum\limits_{\tb} (e_{t} - \lambda) a^\dagger_{\tb} a_{\tb} +
{\frac 1 2}	\sum\limits_{\tb ' s} 
< \tb_1 \tb_2 | V | \tb_3 \tb_4 > 
	a^\dagger_{\tb_1} a^\dagger_{\tb_2} a_{\tb_4} a_{\tb_3}, \label{2}
\end{equation}
where the single particle energies are denoted by $e_t$, the chemical
potential by $\lambda$, and the last term corresponds to the
proton-neutron interaction
\begin{equation}
H_{p ,n} = \sum\limits_{\pb,\pb',\nb,\nb'} 
	< \pb,\nb | V |\pb',\nb' >
	a^\dagger_{\pb} a^\dagger_{\nb} a_{\nb'} a_{\pb'} ~~~.\label{3}
\end{equation}
The subscripts $\tb(t)$ stand for $\pb(p)$(protons) or  $\nb(n)$
(neutrons), being $\tb \equiv t, m_t$, with $t\equiv \{n_t,l_t,j_t\}$ and 
$m_t \equiv m_j$.

Through the Bogoliubov transformation 

\be
\adt = u_t a^{\dagger}_{\bf t} - v_t a_{\bar{\bf t}}, \label{4}
\ee
with $a_{\bar{\bf t}}=(-1)^{t+m_t}a_{t,-mt}$, we get \cite{Pal}
\begin{equation}
H = U + \sum\limits_p 2\omp \e_p \hat\Np   +
	\sum\limits_n 2\omn \e_n \hat\Nn   +
	H_{22} + H_{40} + H_{04} + H_{13} + H_{31} ,\label{hamqp}
\end{equation}
being 
\br
{\hat \N}_{tt'}\equiv {\left[\aadt \aatt\right]^0 \over \sqrt{2\omt}},~
{\hat \N}_{t}\equiv {\hat \N}_{t=t'},~
\omt\equiv {(2j_t+1)\over 2},\nonumber
\er
and where $U$ and the quasiparticle energies $\e_t$ are defined as
\br
U = \sum_t \left[ 2\omt \vt^2(e_t - \lambda) + \sum_{\tt} \sqrt{\omtt  \omt}
\vt^2\vtt^2 F(tt,\tt\tt,0)  - \omt \ut \vt \Delta_t  \right ],
\label{u}  
\er
\br
\e_t = 
\left[e_t -\lambda + \sum_{\tt}\sqrt{\omtt \over
\omt}\vtt^2F(tt,t't',0)\right ]
(\vt^2 - \ut^2) + 2 \Delta_t  \ut \vt\label{7}
\er
being
\br
\Delta_t = -1/2 \sum_{\tt} \sqrt{\omtt \over \omt} \utt \vtt
G(tt,\tt\tt,0),\label{8}
\er
the 'gap' and F, G the usual particle-hole(PH) and particle-particle(PP)
coupled two-particle matrix elements.
The terms $H_{nm}$ in (\ref{hamqp}) destroy $m$ quasiparticles and
creates $n$ quasiparticles, respectively.

To obtain the quasiparticle mean field occupations $v_i$ the ground state
energy $\bra{0} H \ket{0}$ is minimized, {\it i.e.} it is asked that
\cite{Row70,Rin80}
\br
{\partial  \over \partial v} \bra{0} H \ket{0} = 0 \label{dH} ,
\er
subject to the constrictions
\be
\bra{0} \hat N \ket{0} = N, ~~\bra{0} \hat Z \ket{0} = Z ,\label{neq}
\ee
and the normalization $u_i^2 + v_i^2 = 1$, being $\hat N$ and $\hat Z$
the neutron and proton particle number operators respectively.

The standard BCS procedure treats protons and neutrons separately
and in Hamiltonian (\ref{hamqp}) only  pairing interactions, refering 
in general to like particles interacting through the J=0 channel, are
included . 
In this case the ground state energy is just U as defined in Eq.
(\ref{u}) and $\ket{0}$ is the BCS ground state. 
The residual interactions, either between like particles or protons and
neutrons, represented by the terms $H_{nm}$ in (\ref{hamqp}), are usually
included in a second step, most commonly using the QRPA
\cite{Row70,Rin80}.

One of the drawbacks of this procedure is that Eq. (\ref{neq}) is only
enforced for the BCS vacuum. When residual interactions are present
the expectation values of the particle number operators do not coincide
with the actual number. The SCRPA \cite{Sch73,Duk96,Jol80} is  designed 
to overcome this difficulty by solving Eq (\ref{dH}) and (\ref{neq}) 
using the RPA vacuum. 

Full selfconsistency requires to consider the proton-neutron
interaction contribution in the minimization, but the system of equations
which describes this problem is nonlinear and rather complicated, and has
only be implemented in schematic models \cite{Duk96,Pas98,Krm98}. 
In order to perform calculations in realistic Hilbert spaces we will
include in this first step only the like-particles part of the
Hamiltonian, while the ground state $\ket{0}$ will be sensitive to
proton-neutron interaction through the RPA equations, following a
philosophy close to Ref \cite{Bob98}. 
From here on we will refer to this approximation as SRQRPA.
We are absolutely aware that our
treatment is not fully self-consistent. We are just meeting the requirements of Eq.
(\ref{neq}) for the RPA vacuum, and including the modifications in the
mean field due to pn-correlations at the lowest level. However, as is
shown below, even this mild modifications have very important effects in
the observables of the system. 

Following \cite{Jol80}, but including only the modifications of the gap
equations due to use of the RPA vacuum instead of the BCS vacuum, {\it
i.e.} not taking into account the
proton-neutron
 residual interaction explicitly, we
arrive to the modified gap equation

\br
& &\hspace{-1.5cm} 2(\overline{e_t} - \lambda) \vt \ut - (\ut^2 -
\vt^2) \tilde {\Delta}_t = 0,
\label{var1}
\er
where 
\br
\overline{e}_t = e_t + \sum_{\tt}\sqrt{\omtt \over \omt}\vtt^2F(tt,t't',0))
\label{12}
\er
is the single particle energy corrected by the self-energy
and
\br
\tilde{\Delta}_t = -1/2 \sum_{\tt} \sqrt{\omtt \over \omt} \utt \vtt
G(tt,\tt\tt,0)(1 - 2 \bra{0}\hat\Ntt\ket{0}),\label{rgap}
\er
the 'renormalized' gap. Notice that the pairing interaction appears
in the gap equation  renormalized by the presence of proton-neutron residual
interactions, which introduce a finite number of quasiparticles in the RPA
vacuum \cite{Mar98a}. 
For the sake of simplicity we have dropped some higher order terms
in Eq. (\ref{var1}) which come from the interaction between like particles 
connecting the many quasiparticle components of the correlated ground
state. The definition of the renormalized gap (\ref{rgap}) is unique at
this level of approximation \cite{Jol80}. 
This is an important definition to be kept in
mind, because the factor $(1 - 2 \bra{0}\hat\Ntt\ket{0})$ will play a
definite role in suppressing the gap when proton-neutron residual
interactions are large. 

What remains is to couple the renormalized gap problem with the RQRPA
equations.

\section {RQRPA}
\label{sec2}
The nuclear excited states are constructed as \cite{Toi95,Krm96a}

\be 
|\lambda JM\rangle \equiv \Omega^{\dag}(\lambda JM)\ | 0 \rangle,
\label{14}
\ee
\be
\Omega^{\dag}(\lambda JM) = \sum_{pn} \left[ X_{pn}(\lambda J)
 A^{\dag}_{pn}(JM) - Y_{pn}(\lambda J) A_{pn}(\overline{JM})\right],
\label{15}\end{equation}
where
\be
\hat{A}^{\dagger}_{pn}(JM)\equiv 
[ \alpha^\dagger_p \alpha^\dagger_n ]^{JM}
D_{pn}^{-1/2},
~~D_{pn} \equiv \left(1 - \bra{ 0 } \hat\Np + \hat\Nn \ket{ 0 }\right)
,\label{16}
\ee
are the renormalized two-quasiparticle proton-neutron creation operators,
which satisfy 
\be
\bra{ 0 }\left[A_{pn}(JM),A^{\dagger}_{p'n'}(J'M')\right]\ket{ 0 } =
\delta_{pp'}\delta_{nn'} \delta_{JJ'} \delta_{MM'}.\label{17}
\ee
Here $| 0 \rangle$  is the RPA correlated ground state, defined by the
condition
\begin{equation}
\Omega(\lambda JM)| 0 \rangle=0.
\label{vac}
\end{equation}
Each amplitudes $X_{pn}(\lambda J)$ and $Y_{pn}(\lambda J)$ are associated
with the excitation energy $\omega_{\lambda J}$ of the $\lambda$-th
state with angular momentum $J$. They are the eigenvectors and eigenvalues,
respectively, of the RPA equations \cite{Row70,Rin80}

\begin{eqnarray}
\left( 
\begin{array}{ll}  A(J) &  B(J) \\  B^*(J) & A^*(J)\end{array}
\right) 
\left(
\begin{array}{l} X(\lambda J) \\  Y(\lambda J) \end{array}
\right) =
\omega_{\lambda J} \left(
\begin{array}{l} ~ X(\lambda J) \\ - Y(\lambda J) \end{array}
\right),
\label{19} 
\end{eqnarray}
where
\br
A_{pn,p'n'}(J) & = & (\e_p+\e_n)\delta_{pp'}\delta_{nn'}
 + D_{pn}^{1/2}U^{\F}_{pn,p'n'}(J)D_{p'n'}^{1/2}, \nonumber \\
B_{pn,p'n'}(J)& = & D_{pn}^{1/2}U^{\B}_{pn,p'n'}(J)D_{p'n'}^{1/2},
\label{20} 
\er
with
\br
U^{\F}_{pn,p'n'}(J)& = &
G(pn,p'n',JM) (\up\un\upp\unn+\vp\vn\vpp\vnn)\nonumber \\ 
& + & F(pn,p'n',JM) (\up\vn\upp\vnn+\vp\un\vpp\unn), \nonumber \\
U^{\B}_{pn,p'n'}(J)& = &
- G(pn,p'n',JM)(\up\un\vpp\vnn + \vp\vn\upp\unn) \nonumber \\
& + & F(pn,p'n',JM) (\vp\un\upp\vnn+\up\vn\vpp\unn). \label{21} 
\er

The RPA equations (\ref{19}) depend explicitly on the mean field
occupations $v_i$ through the {\bf A} and {\bf B} matrices defined in
(\ref{20}) and (\ref{21}). At the same time, the quasiparticle occupations
$v_i$ depend on the RPA amplitudes $X, Y$, on which depend also 
$| 0 \rangle$, through the renormalized gap
equation (\ref{var1}) and number equations (\ref{neq}).

\section{Expectation values in the correlated RPA vacuum}

Using the quasi-boson approximation 

\be
\left[A_{pn}(JM),A^\dagger_{p'n'}(J'M')\right] \approx 
\bra{ 0 }\left[A_{pn}(JM),A^\dagger_{p'n'}(J'M')\right]\ket{ 0 } =
\delta_{pp'}\delta_{nn'} \delta_{JJ'} \delta_{MM'},\label{25}
\ee
the RPA ground state defined by Eq.(\ref{vac}) can be written as
 
\begin{equation}
| 0 \rangle = N_0 e^{{\hat S}} |BCS\rangle, 
\label{rpavac} 
\end{equation}
with the BCS vacuum defined by the property
\be
\alpha_t |BCS\rangle = 0 , 
\ee
and 
\br
\hat S = \frac{1}{2} \sum_{pnp'n'J} \sqrt{(2J+1)}
C(J)_{pnp'n'}
\left[A^{\dag}_{pn}(J)A^{\dag}_{p'n'}(J)\right]^0 
\label{23}, \\
C(J)_{pnp'n'} = \sum_{\lambda} Y(J)_{pn,\lambda}^*
X(J)^{*^{-1}}_{\lambda,p'n'}. \nonumber
\er

The quasiparticle occupations $\bra{0}\hat\Nt\ket{0}$ evaluated using the 
QRPA vacuum (\ref{rpavac}) have the explicit form \cite{Mar98b}

\br
\bra{ 0 } \hat\Np \ket{ 0 }& = & \sum_{\lambda J n'}{(2J+1) \over
2\omp} D_{pn'} |Y(J)_{pn',\lambda} |^2, \nonumber\\
\bra{ 0 } \hat\Nn \ket{ 0 } & = &\sum_{\lambda J p'}{(2J+1) \over
2\omn} D_{p'n}|Y(J)_{p'n,\lambda} |^2, \label{numqp}
\er

The mean particle numbers of protons and neutrons are \cite{Krm97}
\br
\langle Z \rangle \equiv \bra{0} \hat \Z \ket{0} =  2 \sum_p \omp \vp^2 + 
2 \sum_p \omp (\up^2 - \vp^2) \bra{ 0 } \hat\Np \ket{ 0 },\label{nrpa}\\
\langle N \rangle \equiv \bra{0} \hat \N \ket{0} =  2 \sum_n \omn \vn^2 + 
2 \sum_n \omn (\un^2 - \vn^2) \bra{ 0 } \hat\Nn \ket{ 0 },\nonumber
\er
When the BCS vacuum is used, the second term in each of these expressions
vanish and one is left with the usual number equation. When the residual
interaction is present these terms have a relevant contribution to the
particle number \cite{Mar98a}. They are included in the present work in a
selfconsistent way by modifying the mean field occupations $v_p, v_n$ to
obtain the correct number of particles in the correlated RPA vacuum.

The Fermi transition operators are written in terms of the pair creation
and annihilation operators as \cite{Vog86,Civ87}
\be
\tau^- = \sum_{pn} [v_p u_n [ \alpha^\dagger_p \alpha^\dagger_n ]^{00} +
 u_p v_n [ \alpha_p \alpha_n ]^{00} ]~,
~~\tau^+ = \{ \tau^-\}^\dagger
\ee
which are {\em not} the exact expressions because they are missing the
scattering terms which create a proton (neutron) quasiparticle and
annihilates a neutron (proton) quasiparticle. 

The total strengths $S_\pm$ associated with these transition operators are
\be
S_\pm = \sum_\lambda |\langle \lambda J=0 |\tau^\pm |0 \rangle|^2 .
\ee
The Ikeda sum rule states that, when the exact operators $\tau^\pm$ are
used, and when the set of states $|\lambda \rangle$ is a complete one,
including all the states in the odd-odd nuclei which can be connected with
the ground state $|0 \rangle$ of the even-even nuclei through the
transition operators, then
\be
S_- - S_+ = N - Z \label{ikeda}
\ee

However, in the present case the Fermi transition operators are truncated,
and the strengths difference has a more complicated form \cite{Krm97}
\be
S_- - S_+ = \bra{0} \hat N - \hat Z \ket{0}  -
\sum_{pn} (u_n^2 - v_p^2) \bra{0} \hat \N_n - \hat \N_p \ket{0} .
\ee
The first term is equivalent to $N - Z$ due to the constrictions
(\ref{neq}). The second term gives rise to the violation of the Ikeda
sum rule (\ref{ikeda}). When the BCS vacuum is employed, these term has
no contribution. For this reason the standard pn-QRPA fulfils the Ikeda
sum rule. There are some special cases in which the proton and
neutron quasiparticle occupations in the ground state are equal, and this
term also vanishes, as it was found in  the SO(5) model \cite{Del97} and
in the single mode model of the RQRPA \cite{Krm96a}.
Only when the sum in (\ref{numqp}) is restricted to $J = 0$, forcing
$j_p = j_n$ and $\Omega_p = \Omega_n$, it is true that $\bra{ 0 } 
\hat\N_n - \hat\N_p \ket{ 0 } = 0$ for each level.
In the present case, when a realistic Hilbert space is employed and the
sum over different J's is included in (\ref{numqp}), those occupations
are not equal and the Ikeda sum rule is violated.
It is worth to mention that, while in general the quasiparticle
occupations for protons and neutrons are different, the expectation value
of the total number of quasiparticles $ \langle N_{qp} \rangle
\equiv \sum_\tb \bra{ 0 } \hat\N_t \ket{ 0 }$ is the same, as
can be easily seen making the sum in Eq. (\ref{numqp}).

\section {Results }
\label{sec4}

In the present work we study Fermi beta excitations in $^{76}$Ge. 
We adopt a  $\delta$-type residual interaction already used previously 
\cite{Mar98a},
and our Hilbert space has six single particle energy levels, including 
all the single-particle
orbitals from oscillator shells $3\hbar\omega$ plus
$1g_{9/2}$ and
$1g_{7/2}$ from the $4\hbar\omega$ oscillator shell. They were obtained
using a Coulomb-corrected Wood-Saxon potential. Their numerical values for 
$^{76}$Ge are tabulated in Table 1 of ref.
\cite{Hir90b}. We include $J^\pi= 0^{\pm},..,3^{\pm}$ in the sums in
Eq.(\ref{numqp}).

In order to describe the dependence of the various observables on the
proton-neutron residual interaction we use the parameter
\[
{\rm s}=\frac {v^{pp}_{\pb\nb}} {v^{pair}} \quad, \label{26}
\]
which is the ratio between the coupling constant in the proton-neutron
$J=0$ particle-particle channel and the pairing force constant 
$v^{pair} = (v^{pair}_{\pb\pb}+v^{pair}_{\nb\nb})/2$. 
We compare results  obtained within the usual QRPA with those coming from
the RQRPA (no mean field modification), and the SRQRPA.

In Fig. 1 we show the lowest $J^\pi=0^+$ excitation energies
$\omega$ (in MeV) in $^{76}$As, calculated from the $^{76}$Ge ground
state, for the QRPA (dotted lines), RQRPA (dashed lines) , and
SRQRPA (full lines) approximations, as a function of the residual 
interaction parameter ${\rm s}$. It can be clearly seen that the
excitation energy goes to zero, and collapses around ${\rm s} = 2.5$ for the
QRPA, while in the RQRPA and SRQRPA formalisms the collapse is avoided,
although the excitation energies predicted after the collapse in these two
approaches differ by a factor two.

In Figure 2 the sum rule $S_- - S_+$, normalized to the Ikeda value $N-Z$, 
is presented as a function of ${\rm s}$ with the same convention of Fig. 
1. While in the QRPA the Ikeda sum rule is
always fulfiled, in the other cases it is violated. Notice that,
having the correct number of protons and neutrons in the correlated ground
state, the SRQRPA departs {\em less} from $N-Z$ that the RQRPA, but the
departure is anyway noticeably. It is important to point out that, in the
present calculations, even when ${\rm s}=0$ the other proton-neutron residual
interaction channels, for $J \ne 0$, are present and have a finite value.
For this reason there is a violation of the Ikeda sum rule at ${\rm s}=0$.

To describe in more detail the above mentioned point we present in Figure
3 the total number of quasiparticles as a function of ${\rm s}$. When only $J =
0$
states are included in the sum in Eq. (\ref{numqp}) the number of
quasiparticles goes to zero when ${\rm s}$ is zero. In the other three cases the
residual interaction in the other channels is present and increase this
number.

The fluctuations in the particle number $\Delta N \equiv \sqrt{ 
\langle 0 | (\hat N - N)^2 | 0 \rangle }$
 represent a warning about the region of applicability of the
different QRPA models \cite{Mar98a,Mar98b}. They are shown in Fig. 4a (for
protons) and Fig. 4b (for neutrons) as a function of ${\rm s}$.
It is interesting to note that the pure BCS fluctuations $\Delta N = 2
\sqrt{\sum_t \Omega_t^2 u_t^2 v_t^2}$ diminish when the pn residual
interaction
increases, due to the sharpening in the particle distribution when the
mean field is varied to keep the number of particles in their right
value. In the SRQRPA these fluctuations diminish slightly at first,
dominated by this mean field effect, to increase in a noticeably way after
the QRPA collapse. 
The QRPA results for the total quasiparticle number and the particle
number fluctuations  explode for $s \approx 2.5$ when only the $J^\pi
=0^+$ contribution is included in Eq. (\ref{numqp}).
When all the angular momentum contributions are taken into account
 the $J^\pi =1^+$ component collapses around $s = 2.3$,
and these expectation values diverge. 

The single particle occupation numbers $v_p^2, v_n^2$ are presented
in Figure 5 as a function of the single particle energies. While only six
levels are in use, the curves represent their smoothed values. It can be
seen that both protons and neutrons have their distributions sharpened,
{\it i.e.} the larger ${\rm s}$ the more they resemble a step distribution,
although this effect is more evident for protons. 

The behavior of the pairing energy gap  as a function of ${\rm s}$ is
shown in Fig. 6a (for protons)
and Fig. 6b (for neutrons). As advanced in the previous section, the
pairing gaps, while different for the different single particle levels,
all behave the same way. They show a strong reduction when ${\rm s}$ increases,
down to one half of the original value. This is not a curiosity: the
pairing
energy gap is observable, and any attempt to properly describe the beta
decay using the QRPA extensions must keep the descriptive power of
the simplest formalism, the QRPA, where the pairing constants are
adjusted to reproduce the observed gap.

\section{Conclusions}

A study of the beta decay of $^{76}$Ge using the SRQRPA was
presented. The mean field gap and number equations were introduced and
solved together with the RQRPA equations, but the selfconsistency was only
included to guarantee the good number of particles in the correlated
ground state.

The Ikeda sum rule was studied in some detail. It is  known to be fulfiled
in standard pn-QRPA calculations \cite{Vog86,Civ87}, violated in RQRPA
ones \cite{Hir96} and  recovered when self-consistent RQRPA calculations
are performed in simple models\cite{Krm96a,Del97}. In 
the present work  it was shown that the Ikeda sum rule is violated when a
realistic Hilbert space is used in spite of using the number
selfconsistent QRPA approach. 

The most remarkable phenomenon found in the calculation
was the strong reduction of the pairing gap.
 In the presence of the proton-neutron
residual interaction the mean field changes needed to have the correct
number of particles generate sharper distributions for the
occupations of single particle states,
visible when plotting $v^2$ as a function of the single particle energies
for different values of ${\rm s}$. At the same time, it strongly reduces the
energy gap, which can be as small as a half of the observed value.

Having in mind the various merits of the renormalized
and self-consistent extensions of the quasiparticle random phase
approximation, we conclude that it must be used with extreme caution in
the region where the standard QRPA collapse. Not only the particle number
fluctuations have a clear increase, but observable quantities depart from
their measured values. 

\section{Acknowledgements}

This work was supported in part by Conacyt, M\'exico.

\newpage

\begin{center}
{\bf Figures caption}
\end{center}

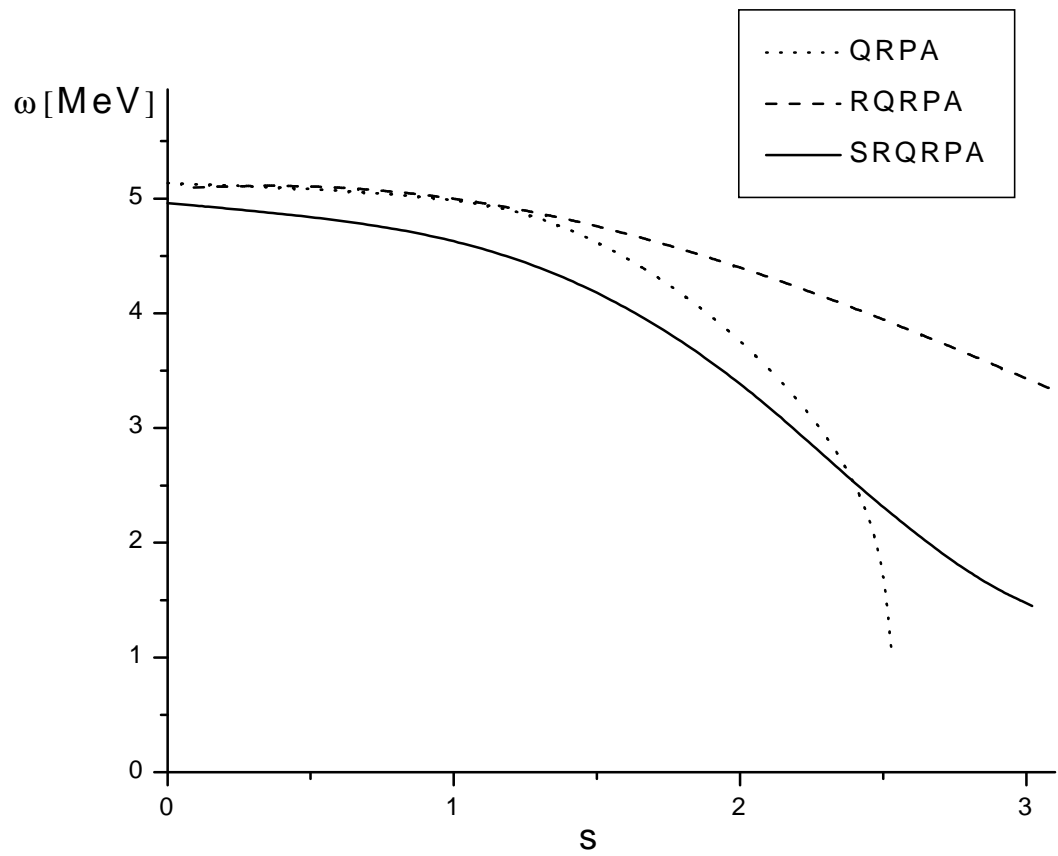
Figure 1. Lowest $J^\pi=0^+$ excitation energies in $^{76}$As, calculated
from the $^{76}$Ge ground state, for the QRPA, RQRPA, and SRQRPA
approximations, as function of the residual interaction parameter ${\rm s}$.

\bigskip

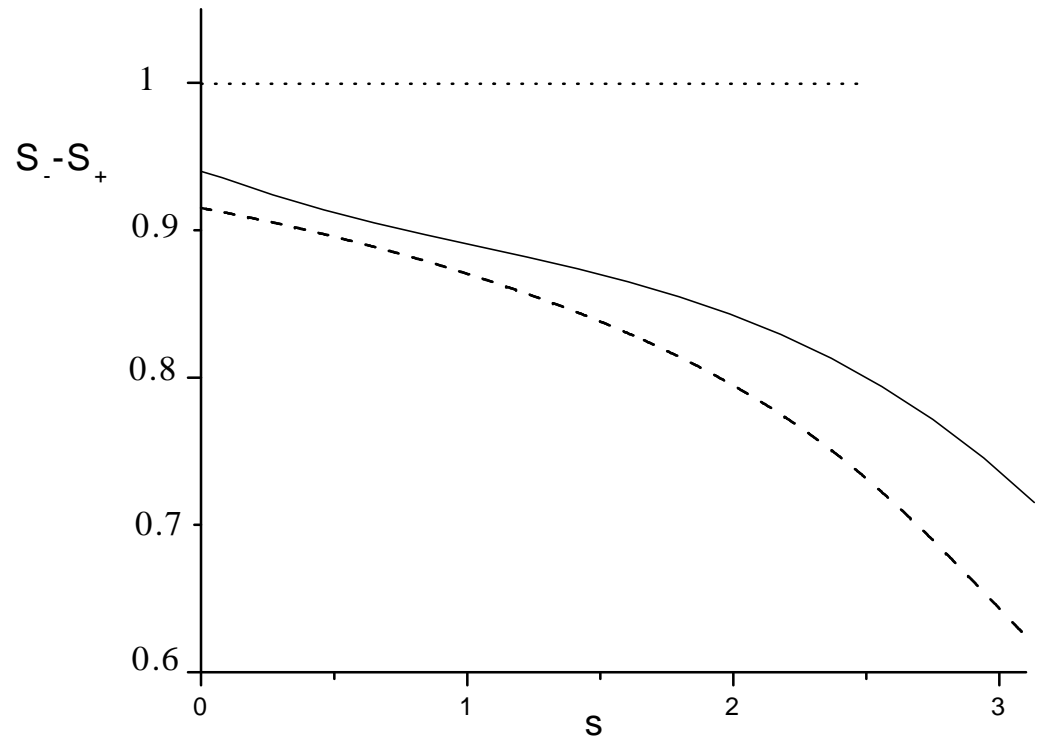
Figure 2. Fermi sum rule $S_- - S_+$, normalized to the Ikeda value $N-Z$, 
as function of ${\rm s}$ with the same line conventions used in fig. 1. 

\bigskip

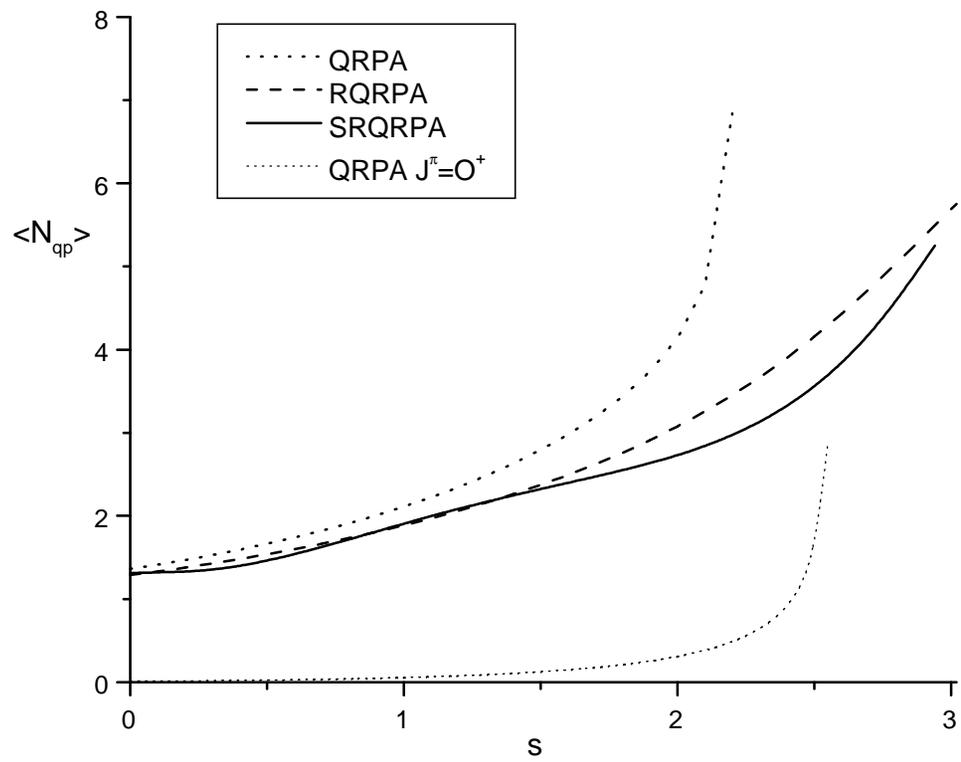
Figure 3. Total quasiparticle number (equal for protons and neutrons)
evolution as function of ${\rm s}$. The result for the QRPA case when
the sums in Eq. (26) are restricted to $J^\pi=0^+$ is also shown.

\bigskip

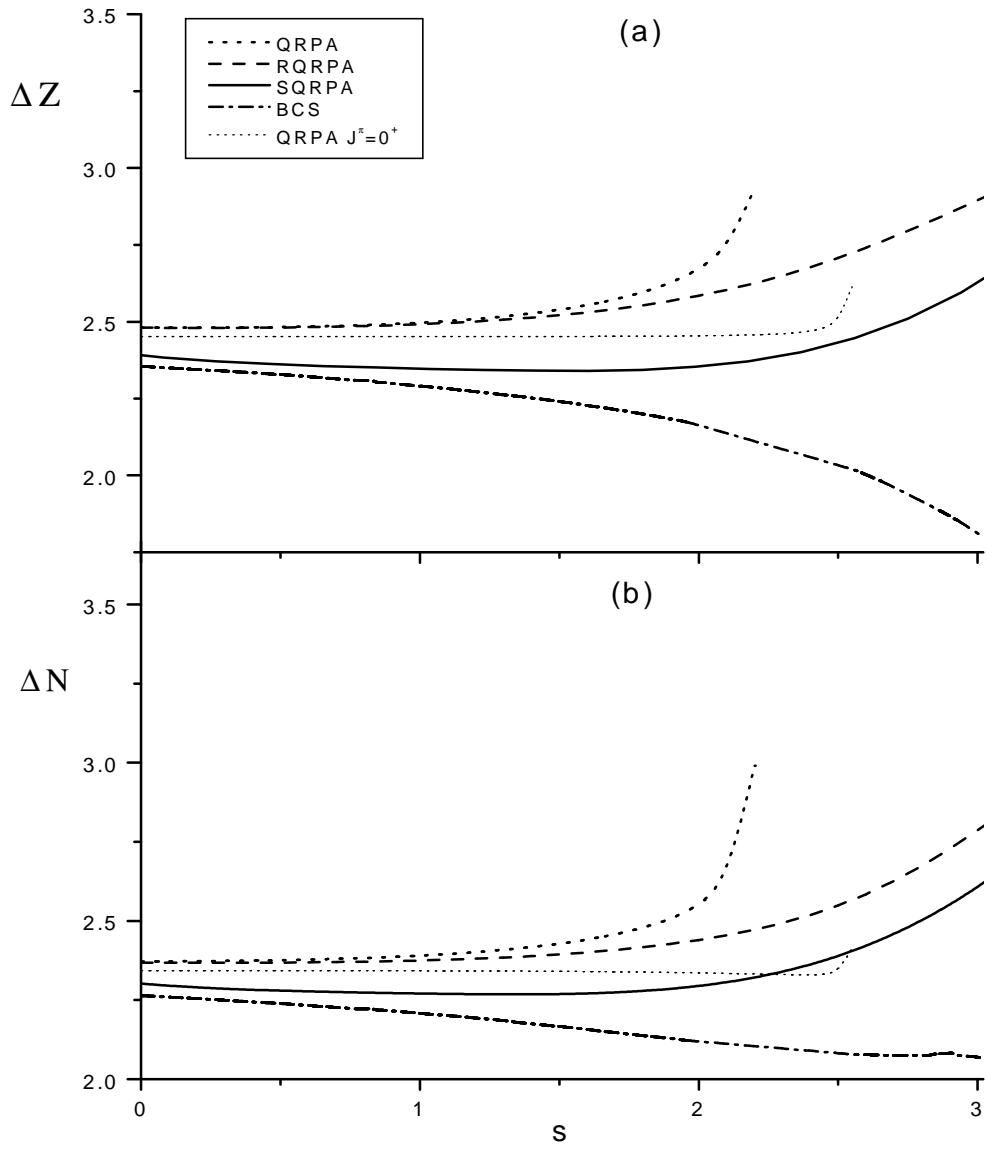
Figure 4. Proton (a) and neutron  (b) number fluctuations obtained
using the BCS, QRPA with only $J^\pi=0^+$ and all the $J$'s in Eq.
(\ref{numqp}), RQRPA and SRQRPA descriptions.

\bigskip

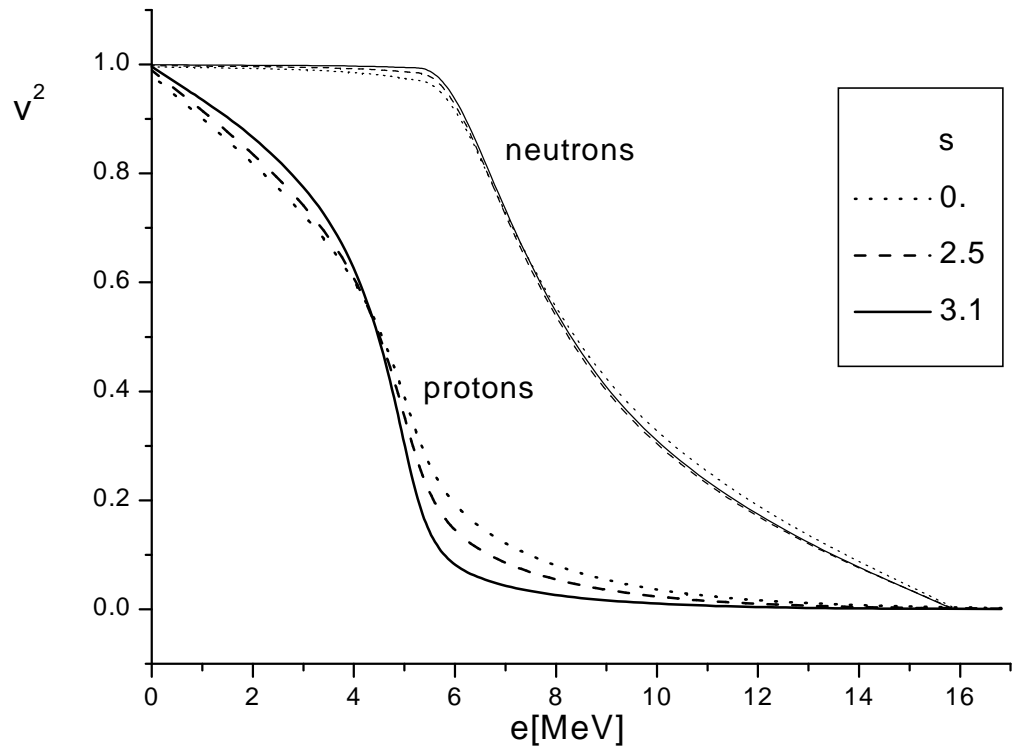
Figure 5. Single particle occupation numbers $v_p^2$ (thick lines), and
$v_n^2$ (thin lines) as a smoothed function of the  single particle
energies $e$, for the values $s = 0., 2.5$, and $3.1$ of the residual
interaction parameter.
\bigskip

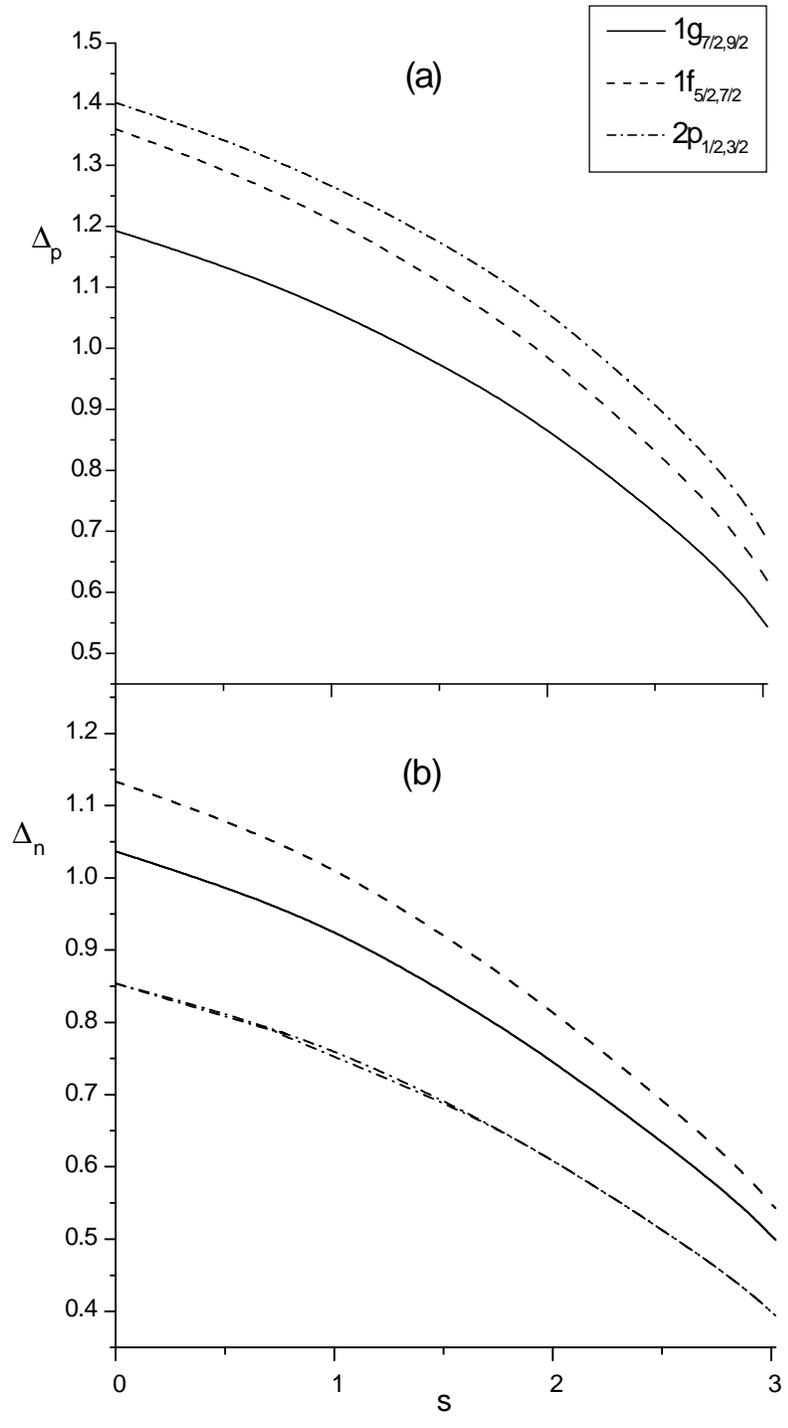
Figure 6.  Evolution  of the pairing energy gap  as a function of ${\rm
s}$ for protons (a) and for neutrons (b). 

\bigskip

\end{document}